\documentclass[aps,prl,twocolumn,superscriptaddress]{revtex4-1}

\usepackage{amsmath,amssymb}
\usepackage{graphicx}
\newcommand{\sfrac}[2]{#1/#2}

\begin{document}

\title{Effects of Fermion Flavor on Exciton Condensation in Double Layer Systems}

\author{J.~Shumway}
\affiliation{Department of Physics, Arizona State University,
 Tempe, AZ 85287}

\author{M.~J.~Gilbert}
\email[]{matthewg@illinois.edu}
\affiliation{Department of Electrical and Computer Engineering,
 University of Illinois, Urbana IL 61801}
\affiliation{Micro and Nano Technology Laboratory, University of Illinois
at Urbana-Champaign, Urbana, IL 61801}

\date{\today}

\begin{abstract}

We use fermionic path integral quantum Monte Carlo to study the effects of fermion flavor on the physical properties of dipolar exciton condensates in double layer systems. We find that by including spin in the system weakens the effective interlayer interaction strength, yet this has very little effect on the Kosterlitz-Thouless transition temperature. We further find that, to obtain the correct description of screening, it is necessary to account for correlation in both the interlayer and intralayer interactions. We show that while the excitonic binding cannot completely surpress screening by additional fermion flavors, their screening effectiveness is reduced leading to a much higher transition temperatures than predicted with large-N analysis.

\end{abstract}


\maketitle

Dipolar fermionic condensates have been a topic of great interest in condensed matter physics for many years. In recent years, double layer systems---two quantum systems separated by a thin tunnel dielectric---have provided a fruitful playground in which to study dipolar superfluidity both experimentally~\cite{Kellogg:2004, Tutuc:2004, Snoke:2002, TiemannC:2008} and theoretically~\cite{Wen:1992, Fertig:1989, Moon:1995, Park:2006}. Interest in dipolar superfluids has received increased due in large part to the prediction of dipolar superfluid behavior at or above room temperature in double layer graphene~\cite{Min:2008,Gilbert:2009,Zhang:2008}. This is uniquely possible in graphene due to the symmetric linear band-structure and ability to sustain large carrier concentrations in two closely-spaced layers. Yet, this prediction is not without significant controversy. As superfluidity is predicted to occur in the double layer graphene system outside of the quantum Hall regime, additional fermion flavors---beyond the top or bottom layer freedom---may participate in the phase transition. Theoretical disagreements over the Kosterlitz-Thouless transition temperature ($T_{KT}$) arise from differing assumptions about the importance of these extra flavors for screening in dipolar exciton condensates. In the works predicting a high transition temperature of $T_{KT}\approx 0.1$ $T_F$ (where $T_F$ is the system Fermi temperature) the fermionic degrees of freedom in the system were taken to be strongly correlated  in a gapped condensate phase and could not screen enough to significantly lower $T_{KT}$. Other works  that predict a low transition temperature of $T_{KT}\approx10^{-7}$ $T_F$ assume that screening from additional degrees of freedom add independenty to effectively screen out the the interlayer interaction~\cite{Kharitonov:2008, Kharitonov:2010}. While experiment will be the ultimate arbiter of the value of $T_{KT}$, many-body theoretical approaches will play a significant role in understanding the nature of the of the phase transition.

In this Letter, we use fermionic path integral quantum Monte Carlo (PIMC)~\cite{Pollock:1984, Ceperley:1995} to elucidate the role of screening in exciton condensates formed in symmetric electron-hole double layer systems. We show that increased fermion flavor does increase the screening in exciton condensates formed in symmetric double layer system. However, the role of screening is not as dramatic as predicted in previous analytic work, because of strong excitonic pairing and other correlations. We compare the static correlations and dynamic response functions of symmetric electron-hole double layer systems with different number of fermionic flavors. We use a symmetric model of electrons in holes in two-dimensional sheets, separated by a $d=0.5$~nm layer of insulating SiO$_2$. Through analysis topological winding numbers and pair-correlation functions, we show that when the spin degree of freedom is included in our simulations, the transition temperature for our system drops roughly with $T_F$, from $T^{ns}_{KT}\approx 580$~K~$\approx 0.19~T_F$ to $T^{sp}_{KT} \approx 370$~K$\approx 0.24~T_F$, indicating that the increased number of fermion flavors participating in double layer systems are not completely screened by the by the condensate. Furthermore, we show that the inclusion of intralayer interactions with spin-\sfrac{1}{2} fermions, in combination with the high layer carrier concentrations and attractive exciton-exciton interaction, allow the system to form biexcitons. Additionally, we use dynamic density-density response function, collected within the PIMC framework, to demonstrate that when the spin degree of freedom is included, the polarizability of the system actually decreases corresponding to a complicated interplay between Pauli exclusion forcing identical carriers to avoid one another, increasing screening, and that neutral excitons do not screen charge effectively.

The Hamiltonian for our system includes the kinetic energy of the quasiparticles and Coulomb interactions,
\begin{equation}
\begin{split}
H  = &\sum_{i=1}^{N_e} \frac{p_{i,e}^2}{2m^*}
+ \sum_{i=1}^{N_h} \frac{p_{i,h}^2}{2m^*}
+\sum_{i<j} \frac{e^2}{\epsilon|\mathbf{r}_{i,e}-\mathbf{r}_{j,e}|} \\
+&\sum_{i<j} \frac{e^2}{\epsilon|\mathbf{r}_{i,h}-\mathbf{r}_{j,h}|}
-\sum_{i,j} \frac{e^2}{\epsilon
\sqrt{|\mathbf{r}_{i,h}-\mathbf{r}_{j,h}|^2 + d^2}}.
\end{split}\label{eq:Hamiltonian}
\end{equation}
All quasiparticles have equal mass of $m=0.09 m_e$, so that the Fermi veloctiy approximates the velocity of quasiparticle in graphene~\cite{Gilbert:2009}. We take the dielectric constant of the tunnel barrier to be $\epsilon_r=3.9$ corresponding to SiO$_2$. In this work we separately consider the cases of spinless and spin-\sfrac{1}{2} quasiparticles, corresponding to different numbers of fermion flavors: $N_f=2$ when the carriers only have a layer degree of freedom, and $N_f=4$ when the carriers have both a layer and spin degree of freedom. Each of the layers in our system is assumed to be $20$ nm~$\times~20$ nm with periodic boundary conditions  the plane of the layers. We use $N=40$ quasiparticles in our simulations consisting of $20$ electron and $20$ holes per layer.  The number of electrons and holes corresponds to carrier densities of $n_e=n_h=5\times10^{12}$ cm$^{-2}$ and places the system in the exciton condensate regime~\cite{Min:2008,Gilbert:2009,Zhang:2008}. When simulations with spin are performed, we use $10$ spin-up and $10$ spin-down particles per layer. Since the Hamiltonian does not have any spin-dependent terms, the spin only enters into the calculation when considering fermion anti-symmetry upon particle exchange. The Coulomb interactions are handled within the pair approximation, and the path is discretized with a time step $\hbar\,\Delta\tau \approx 3$~Ha$^{-1}$; for example, 256 slices are used to represent the path when $T=400$~K. To circumvent the well-known fermion sign problem in our simulations, we use a ground-state fixed node approximation~\cite{Anderson:1976, Ceperley:1992}, as used in three-dimensional excitonic BEC studies~\cite{Shumway:2000}. At each imaginary time slice, we require that the electron and hole coordinates lead to a postive Slater determinant, $\det|\phi(r_{e,i}-r_{h,j})|$ where $\phi = e^{-r/a}$ is the typical BCS mean-field pairing wavefunction typically used in quantum Monte Carlo simulations of exciton condensation in symmetric electron-hole systems~\cite{Zhu:1995, DePalo:2002}. For our simulations, we have used an exciton pairing radius, $a$, of $2.3$ nm but have performed simulations containing $a=1$~nm to $10$~nm and find our results to be insensitive to this parameter.
%
\begin{figure}[t]
\includegraphics[width=\columnwidth]{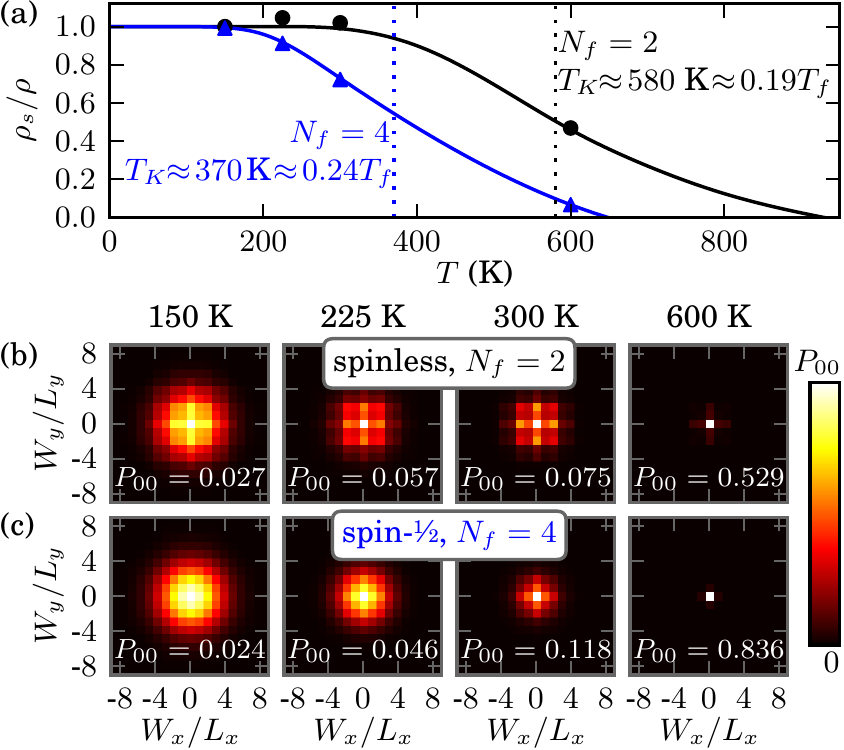}
\caption{(color online) (a) Superfluid fraction for spinless (black circles, $N_f=2$) and spin-\sfrac{1}{2} (blue triangles, $N_f=4$) symmetric double-layer electron-hole condensate, calculated from PIMC simulations of 40 electron-hole pairs. Estimates of $T_{KT}$ are made where the superfluid fraction drops to $1/2$, and lines are a guide to the eye. 
Normalized historgrams of topological winding distributions, which measure superfluid density, Eq.~(\ref{eq:sfd}), for the (b) spinless and (c) spin-\sfrac{1}{2} case.\label{fig:sffrac}}
\end{figure}

In Fig. \ref{fig:sffrac}~(a), we show our calculated superfluid fraction as a function of the system temperature. Our calculations reveal a suppressed $T_{KT}$ when the additional spin degree of freedom is included. Superfluid fraction is an excellent way to understand phase transitions in systems as it saturates at $1$ for low temperatures and asymptotes to $0$ beyond the phase boundary. We denote the $T_{KT}$ as the temperature where the superfluid fraction has dropped to $0.5$ to be consistent with previous work~\cite{Pollock:1987}. We see a clear drop in $T_{KT}$ from $T_{KT}\approx 580$ K in the spinless case to $T_{KT}\approx 300$ K in the spin-\sfrac{1}{2} case, illustrating the effect of increased fermion flavor on the phase transition. In fact, this drop in transition temperature may be understood without resorting to screening arguments. One would expect a decrease in $T_{KT}$ simply from reduced quantum degeneracy as more fermion flavors are added. In an ideal Bose gas, the magnitude of the transition temperature is determined by the condition $n\lambda(T_{KT})^2\sim 1$, where $\lambda(T)\sim T^{-1/2}$ is the thermal deBroglie wavelength and $n$ is the density of identical Bosons. As fermions pair into bosons, there must be at least $N_f$ distinguishable species of bosons, decrease the density $n$ and hence, the transition temperature $T_{KT}$ by $1/N_f$. This effect of reduced quantum degeneracy degeneracy can also be seen by comparing $T_{KT}$ to the Fermi temperature (at our density,
$T_F^{ns}= 3086$~K and $T_F^{sp}= 1543$~K); in each case our estimated $T_{KT}$ is around four or five times smaller than $T_F$.

The calculated superfluid fraction is estimated by the presence of permuting paths that wind around the periodic box~\cite{Ceperley:1995}. In two dimensions, the superfluid fraction is given by,
\begin{equation}\label{eq:sfd}
\frac{\rho_s}{\rho}
= \frac{mk_BT}{\hbar^2 (N_e+N_h)} \langle W_x^2 + W_y^2\rangle.
\end{equation}
In Eq. (\ref{eq:sfd}), $W_x$ and $W_y$ are the topological windings (which are integer multiples of the supercell dimensions) corresponding to the different path configurations. Note that for excitonic condensates, we use the number-coupled winding, where $W_x$ and $W_y$  each include a sum over individual windings of electron and hole quasi-particles.

In Fig. \ref{eq:sfd}, the bottom two rows, (a) and (b) compare the exciton winding probabilities for the spinless (a) and spin-\sfrac{1}{2} (b) for several temperatures. At $T = 150$ K, we see similar behavior for both systems, as each system is nearly all superfluid. Careful examination shows the histogram for the spinless case have extra even-odd structure. In an exciton, electrons pair with the holes and wind as a composite particle giving rise to peaks in the winding histogram at even integers. Thus showing clear evidence the system is an exciton condensate. The even-odd effect is more washed out in the spin-\sfrac{1}{2} state, which we attribute to the relatively small number, ten, of identical fermions. Larger simulations should sharpen this even-odd signature of an excitonic condensate. Note that biexcitons would have strong peaks at multiples of four; we do not see evidence of a biexcitonic condensate in our simulations.

\begin{figure}[tbp]
\includegraphics[width=\columnwidth]{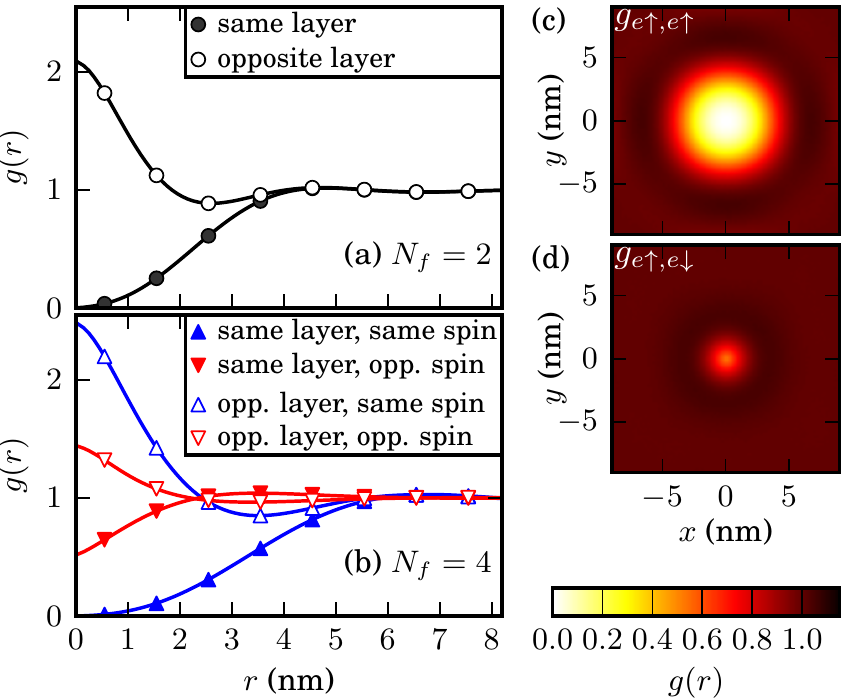}
\caption{(color online)
Radial pair correlation functions for (a) spinless, $N_f=2$, and (b) spin-\sfrac{1}{2}, $N_f=4$
cases.
(a) Intralayer exchange-correlation hole for the same spin 
species, $g^{\uparrow\uparrow}({\bf r}_{ij})$. (d) Intralayer correlation hole 
for opposite spin species, $g^{\uparrow\downarrow}({\bf r}_{ij})$. \label{fig:pcf}}
\end{figure}

Next we examine real-space static correlation among the quasiparticles, due to Coulomb interactions and Pauli exclusion. We have collected the pair-correlation functions $g(r)$, which are  easily calculated in PIMC by binning the equal-time pair distributions in a histogram. In contrast to other QMC methods, such as variational or diffusion Monte Carlo, there is no bias from a trail wavefunction, thus PIMC gives essentially exact pair correlations, aside from errors associated with the fixed-node approximation. In  Fig.~\ref{fig:pcf} we show our calculated pair correlation functions at $T = 150$ K. We find the presence of an exchange hole in between identical fermions, which is larger in the $N_f=4$ case simply because the density of particles of a particular flavor decreases as $1/N_f$. We find that there is a strong, attractive, interlayer correlation $g^{eh}$, with a peak at small separations that is coincident with the formation of indirectly bounds excitons with holes in the top layer sitting directly above the electrons in the bottom layer. This pairing of identical spin states in the $N_f=4$ case may be an artifact of
our nodal model, where we have built or nodal wavefunction from a Slater determinant of 
orbitals pairing like spins only.

Interestingly, we find a small correlation hole for the $N_f=4$ case, arising almost equally from the electron and hole of opposite spin: the particle in the same layer (with the same charge) is slightly repelled, while the particle in the opposite layer (with the opposite charge) is similarly attracted. This is a clear demonstration extra fermion flavors correlate among each other, even when they do not bind as condensed excitons.
This is more dramatically illustrated in Fig. \ref{fig:pcf} (c) and (d). In Fig. \ref{fig:pcf}, we plot the intralayer pair correlation in 2D corresponding to $g_{sp}^{\uparrow\uparrow}$ and, as expected, we see a large correlation hole where Pauli exclusion and Coulomb repulsion forces other identical particles apart. However, when we include spin-\sfrac{1}{2} particles and examine the intralayer correlation of $g_{sp}^{\uparrow\downarrow}$ in Fig. \ref{fig:pcf} (c), we see a much weaker correlation hole. Note that we again do not see evidence of biexciton formation, which would appear as a positive correlation of like-charged, opposite-spin particles in the same layer~\cite{Shumway:2000}.

\begin{figure}[tbp]
\includegraphics[width=\columnwidth]{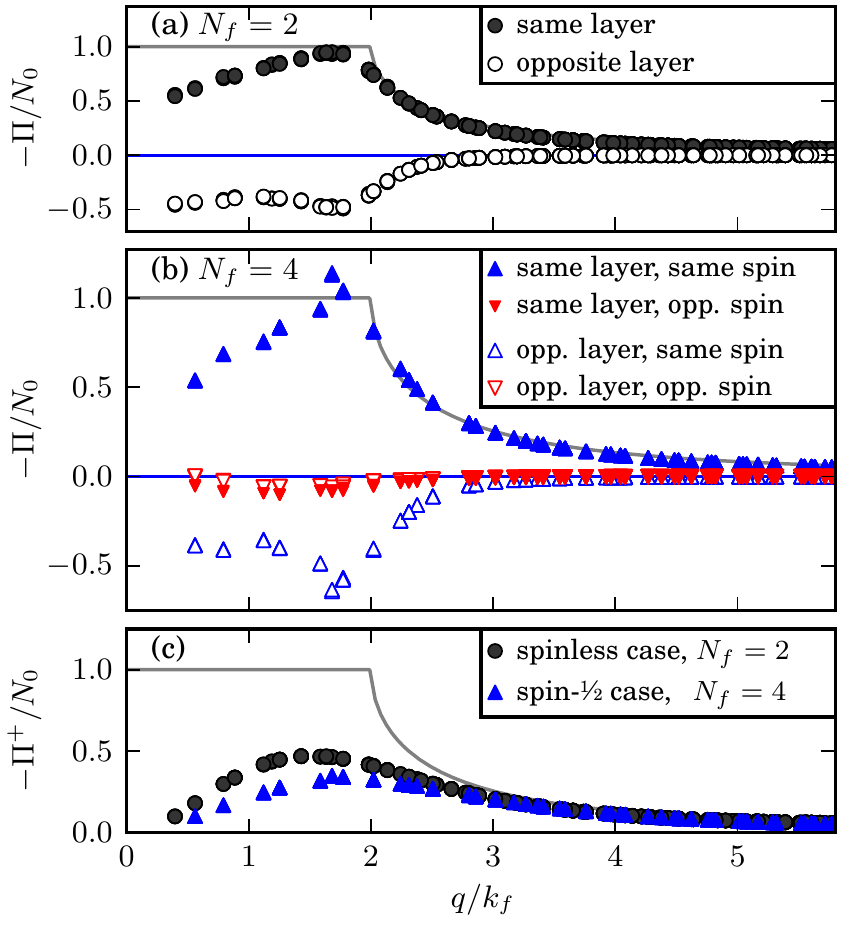}
\caption{(a) Interlayer and intralayer polarizabilities for $N_f=2$ versus the perturbation wavevector normalized by the Fermi wave vector. (b) Interlayer and intralayer polarizabilities for $N_f=4$ normalized by the Fermi wave vector. (c) Sum of the interlayer and intralayer polarizabilites for $N_f=2$ and $N_f=4$ normalized by the Fermi wave vector. In each plot the polarizabilities are plotted against the Lindhard response function (grey line) in 2D and have been normalized by the density of states at the Fermi energy in 2D, $N_0$.}
\label{fig:polar}
\end{figure}

We do not see a large drop in $T_{KT}$  as predicted in Refs.~\onlinecite{Kharitonov:2008, Kharitonov:2010}. In an effort to understand this, we explore the dynamic correlation functions, namely the polarizability. In Refs.~\onlinecite{Kharitonov:2008,Kharitonov:2010},
the large-$N_f$ approximation strongly resembles the random phase approximation (RPA),
and hence the screening they find is essentially the Lindhard function, with suppression of
screening at very small $q$ due a BCS-like excitonic condensate.
In Fig.~\ref{fig:polar}, we show that our calculated polarizabilities are considerably different.
To collect the dynamic correlation functions we sample the polarization operator between different fermion flavors
$\gamma$ and $\gamma'$,
\begin{equation}\label{eq:polarization}
\Pi_{\gamma\gamma'}(\mathbf{q},i\omega_n) = -\frac{q_\gamma q_{\gamma'}}{V\hbar}
\int_0^{\beta\hbar}
\langle T_\tau\; n_{\mathbf{q},\gamma}(\tau)n_{-\mathbf{q},\gamma'}(0)
\rangle\;d\tau,
\end{equation}
where $n_{\mathbf{q},\gamma}=\sum_{i=1}^{N_\gamma} e^{i\mathbf{q}\cdot\hat{\mathbf{r}}_i}$
is the density operator for fermion flavor $\gamma$.

In Fig.~\ref{fig:polar} we show the $\Pi_{\gamma,\gamma'}(\mathbf{q},0)$ between different fermion flavors normalized by the 2D density of states as a function of wavelength normalized by the Fermi wave vector at a system temperature of $T=150$~K. Here we find the polarization response from identical particles in $N_f=2$ and $N_f=4$ track the Lindhard function very well in the short wavelength limit, past $q>2k_f$, as one would expect from Fermi liquid theory. 
 Interestingly, we find similar behavior for both $N_f=2$ and $N_f=4$ in response to long wavelength excitations ($q \rightarrow 0$). This is expected as we are clearly in the condensate regime in which all carriers should be paired into excitons and, therefore, unable to respond to perturbations. The response between  particles in different layers or of different spin is negative, indicating that these correlations suppress screening from the independent particle, RPA response. The response of the exciton is quite strong, which the opposite-spin flavors have a weak negative response, Fig.~\ref{fig:polar}(b). We have calculated the polarizabilities at temperatures both above and below T$_{KT}$,  and found very little difference between the different temperature traces for both the $N_f=2$ and $N_f=4$ cases, indicating that the screening behavior is dominated by pre-formed excitons, with exist above $T_{KT}$.

However, Fig.~\ref{fig:polar}(c) shows that the sum of inter- and intra-layer polarizabilities for $N_f=4$ is smaller than that of $N_f=2$. While the large-N calculation finds enhanced screening with more fermion flavors $N_f$ which grows as $N$, but we find a weaker  trend. The reason for this difference is excitonic binding and correlation between different. In the large-N expansion, the starting point is RPA like screening. Because each flavor can screen independently, it has been argue that the Coulomb interaction because too weak to form pairs, except for weak BCS pairing of states near the Fermi surface when the temperature is below T$_{KT}$. In our simulations, our model has strong excitonic pairing. The pairing of charge into neutral excitons suppresses their ability to screen. This excitonic pairing and correlation between quasiparticles of different spin is not present in the large-$N_f$ approximation's perturbative expansion.

In conclusion, we have analyzed the static and dynamic response functions obtained from fermionic PIMC quantum Monte Carlo simulations of symmetric electron-hole double layer systems which are spaced a distance of $d = 0.5$ nm. We find that the addition of extra fermion flavors to the system reduces T$_{KT}$ because the additional flavors are not screened out by the excitons in the condensate thereby weakening the interlayer interaction strength which drives the transition from Fermi liquid to exciton condensate. However, we show that while fermion flavors cannot be ignored the drop in T$_{KT}$ is not as large as predicted by the large-N expansion calculations. Our analysis of polarizability shows that the strong excitonic pairing in our model suppresses screening allowing a higher transition temperature.

\begin{acknowledgments}
We wish to acknowledge M. Y. Alaoui Lamrani and Z. Estrada for help running the simulations. MJG would like to thank D. M. Ceperley for insightful discussions. MJG is supported by the Army Research Office (ARO). JS is supported by the Semiconductor Research Corporation (SRC) Nanoelectronics Research Initiative (NRI) South West Academy of Nanoelectronics (SWAN). Computer simulations used the National Science Foundation TeraGrid and the ASU Advance Computing Center (A2C2).\end{acknowledgments}

\bibliography{bilayer}

\end{document}